# Evacuation simulation considering action of the guard in an artificial attack

Chang-kun Chen(陈长坤)[1] and Yun-he Tong（童蕴贺）[1†]

[1] *Institute of Disaster Prevention Science & Safety Technology, Central South University, Changsha, 410075, P.R. China*

To investigate the evacuation behaviors of pedestrians considering action of the guard and develop an effective evacuation strategy in the artificial attack, an extended floor field model was proposed. In this model, the attacker's assault on pedestrians, the death of pedestrians and the guard's capture were involved simultaneously. An alternative evacuation strategy which can largely reduce the number of death was developed and effects of several key parameters such as the deterrence radius and capture distance on evacuation dynamics were studied. Results show that congestion near the exit has dual effects. More specially, the guard could catch all attackers in a short time because the attackers would have more concentrated distribution, but more casualties would happen because pedestrians are hard to escape the attack due to congestion. In contrast, when pedestrians have more preference of approaching the guard, although the guard would take more time to capture the attackers result from the dispersion of attackers, the death toll would decrease. One of the reason is the dispersal of the crowd and the decrease in congestion would be beneficial for escape. Another is the attackers would be killed before launching the attack to the people those are around the guard, in other words, the guard would protect a large number of pedestrians from being killed. Moreover, increasing capture distance of the guard can effectively reduce the casualties and the catch time. As the deterrence radius reflecting the tendency of escaping from the guard for attackers rises, it would become more difficult for the guard

*Project supported by National Key Research and Development Program of China (No.2017YFC0804900) and National Natural Science Foundation of China (Nos.71790613 and 51534008).

† Corresponding author. E-mail: ohmytong@163.com

to catch the attackers and more casualties are caused. However, when the deterrence radius reaches a certain level, the number of deaths would be reduced because the attackers would prefer to stay as far away as possible from instead of the position where they could attack more people.

1. Introduction

In recent years, pedestrian evacuation under the artificial attacks has attracted much attention due to many realistic events. For example, the knife attack carried out by a 35-year-old man with a personal grievance in a busy shopping mall in Beijing on February 11, 2018, and left 1 dead, 12 injured, and the deadly mass knife attack by knife-wielding men at a railway station in Kunming in south-west China on March 1, 2014 and left at least 29 dead, more than 130 injured. Due to the threat of artificial attacks, it is necessary to understand the pedestrian dynamics in this situation for developing effective evacuation schemes.

During the last decades, various simulation models have been established to investigate pedestrian evacuation processes [1], and there are two fundamentally different ways of representing people in these models, namely macroscopic models and microscopic models [2]. According to the macroscopic models, pedestrians are represented as an analogy to fluid flow with a specific density which corresponds to people density and velocity [3]. In contrast, each pedestrian in microscopic models would be treated as a self-driving particle with certain properties [4]. Therefore, these models could consider the heterogeneities of pedestrians, which make it more similar to reality and become the most common way of modeling pedestrian dynamics [5]. Particularly, floor field model [6], one of the most important microscopic model, is a well-studied pedestrian model using cellular automata[7]. Due to its flexibility and extensibility[8], floor field model is extensively used and could successfully reproduce realistic pedestrian behavior and self-organization encounter in pedestrians dynamics, such as clogging[9], exit selection strategy [10], leading [11] and group behavior [12] conflicts at the exit[13], crowd flow through multiple bottlenecks[14], oscillation at the bottleneck[15] [16].

Pedestrian flow exhibits variable patterns of behavior in the artificial attack scenario, and the most important reason is the complex interactions involved in this situation, such as pedestrian-to-pedestrian, pedestrian-to-environment, attacker-to-pedestrian and pedestrian-to-attacker interactions. Each interaction can be described as a floor filed in floor field model where individuals make their decision according to the so-called transition probabilities modified by different floor fields. Chen [17] studied pedestrian dynamics by an extended floor field model and found the rolling behavior and along-the-wall motion of the crowd with aggravating extent of the impact of attackers on pedestrians. Li [18] proposes a three-stage model to reproduce a series of complex behaviors and decision-making processes at the onset of an attack, and the impact of the terrorist attack on pedestrian dynamics has been well-studied. Liu [19] developed a social force model to study the crowd evacuation when a terrorist attack occurs in the public place and the effects of the initial positions of terrorists, the terrorist number and the emergency exit choice strategy on crowd evacuation have been studied.

However, fewer researchers focus on the pedestrian evacuation involves the attacker and the guard simultaneously. In reality, there are always guards with protection function in public areas and the guard would have a positive effect on pedestrian but a negative effect on the attackers, which make pedestrian dynamics different and more complicated. It should be studied for reducing the death toll and developing effective evacuation strategy.

In this paper, an extended floor field model was proposed to investigate evacuation behaviors of pedestrians in an artificial attack considering the guard, and the movement of the pedestrian, guard and attacker, the attacker's assault, and the guard's capture were involved simultaneously. The rest of this paper is organized as follows: the proposed model considering the interactions between the attacker, guard and pedestrian is introduced in section 2. In section 3, the comparison of two evacuation strategies and the effects of several key parameters on pedestrian evacuation are discussed. In section 4, the conclusion is made and the paper is closed.

## 2 Model description

In the proposed model, the space is represented by two-dimensional foursquare cells. Each cell is an identical square of 0.4 m×0.4m [20] and can either be empty or occupied by an obstacle, a pedestrian, a guard or an attacker. In each discrete time step, they could move one step by corresponding principles. More specially, the attackers would aim to kill more people so they would be attracted by the crowd and move towards the desired direction according to the calculated attractive force. However, once the distance from the guard is small enough, the attackers have to run far away from the guard as possible to avoid being caught. On the other hand, the guard would chase the nearest attacker and could kill them in a capture distance by the prey-predator model. Meanwhile, the pedestrian would be killed with a certain probability when attacked by the attacker and the movement of pedestrians would be based on the extended floor field model where individuals make their decision according to the so-called transition probabilities modified by the exit, attack threat and the guard floor field. Three different kinds of actions are involved in this model and their detailed expressions are modeled as follows.

### 2.1 Action of the attacker

In general, the artificial attackers aim to attack more people and create panic as much as possible thus the attacker does not have clear targets and are assumed to be attracted by pedestrians and move towards a direction of the larger population. However, sometimes the attacker might run away from the guard to avoid being attack if the guard is getting very close the attacker. Therefore, a new parameter $R_D$ named deterrence radius was introduced to represent the critical distance that the attackers have to escape from the guard, which can be viewed as a measure of the tendency of attackers to avoid the guard while they are chasing the crowd. In other words, the attackers would run away from the guard when the distance between the attackers and the guard is less than deterrence radius, otherwise, the attackers would chase and attack the crowd. It should be noted that the attacker would also attack the pedestrian during this escape process. Accordingly, either the intensity to move towards the crowd or the repulsion between the attacker and the guard was imposed on the attacker, and an analogical formulation taking reference of the physical mechanics was introduced. And a detailed expression for the above forces is modeled as Fig.1 shown.

$$\vec{f} = \begin{cases} \dfrac{\sum_{n=1}^{n}\vec{f_{p_n}}}{\left|\sum_{n=1}^{n}\vec{f_{p_n}}\right|} & d_{ag} > R_D \\ \dfrac{\vec{f_{ag}}}{\left|\vec{f_{ag}}\right|} & d_{ag} \leq R_D \end{cases} \tag{1}$$

where $\vec{f}$ is the resultant force on the attacker. $n$ is the number of pedestrians within the attacker's sighting range. $\vec{f_{p_n}}$ is the single attractive force of the $n^{th}$ pedestrian to the attacker. $\sum_{n=1}^{n}\vec{f_{p_n}}$ is the intensity to move towards the crowd. $\vec{f_{ag}}$ is the repulsion between the attacker and the guard and $d_{ag}$ is the distance between the attacker and the guard. $R_D$ is the deterrence radius.

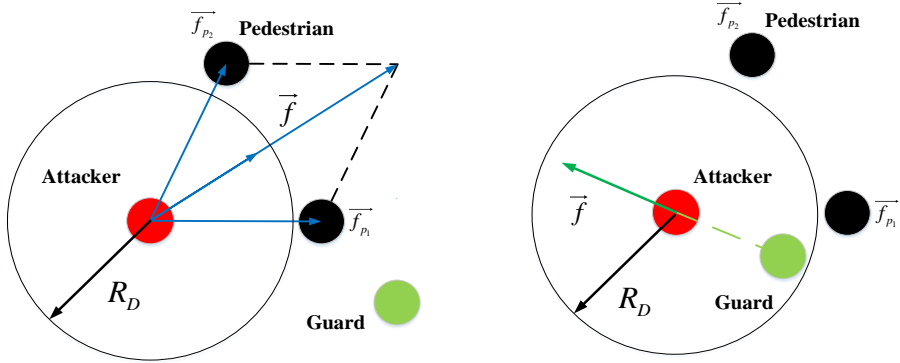

Figure 1 Diagram of the force to the attacker when $n=2$.

In Fig.1, the red circle represents the attacker, the black is the pedestrian, and the green is the guard. Each pedestrian has a single attractive force to the attacker. $\vec{f}$ denotes the force to the attacker and $R_D$ is the deterrence radius. In the first case, $\vec{f}$ is calculated by the attractive resultant force from the crowd because the distance between the attacker and the guard is bigger than the deterrence radius and the attacker would chase the crowd. While in the other case, $\vec{f}$ is calculated by the repulsion from the guard which results from that the distance between the attacker and the guard is smaller than the deterrence radius and the attacker has to get away from the attacker to avoid being attacked.

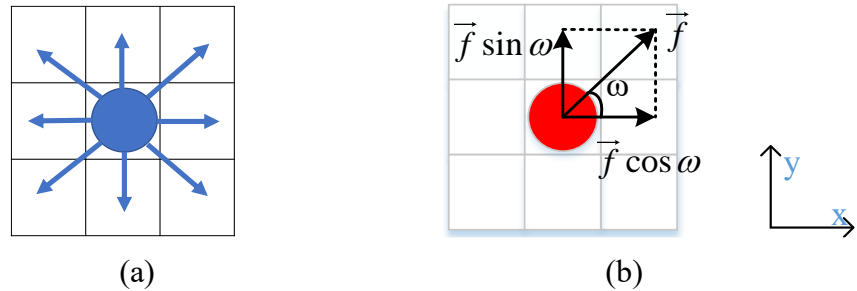

Figure.2 Possible movement directions of the attacker(a) and component force of $\vec{f}$ at X-axis and Y-axis.

As shown in Fig.2(a), for the attacker, except keeping unmoved, there are eight possible movable directions at the next time step, but which direction the cell will move to depends on the resultant force figured out by Eq. (1). As illustrated in Fig.2(b), $\vec{f}$ denotes the resultant force on the attacker hypothetically, and the component force of $\vec{f}$ at X-axis and Y-axis could be expressed as $\vec{f}\cos\omega$ and $\vec{f}\sin\omega$ respectively, and the movable direction for the attacker at the next time step could be determined by the relative magnitude of $\vec{f}\cos\omega$ and $\vec{f}\sin\omega$. When pedestrian $n$ is assumed to be located at the position *(i, j)*, the relationship between the relative magnitude of $\vec{f}\cos\omega$ as well as $\vec{f}\sin\omega$ and the rule of movable direction for pedestrian $n$ at the next step were tabulated in Table 1.

The assault of the attackers would be considered this model. The attacker will launch attacks on the pedestrian when the distance between the attacker and the pedestrian is small enough, and the pedestrian will be killed with a certain probability. And the attacker could just attack one time at each time step and the probability of the pedestrian being killed is set as 0.7 [19] in this model.

## 2.2 Action of the guard

The interaction between the guard and attackers is a pursuit-and-evasion problem and can be modeled by the prey-predator model [21], which is extensively used to study collective motion of living organisms and could reproduce many self-organization phenomena [22]. In this model, the guard can be regard as the predator and the attacker is the prey, and the guard would be assumed to be have more force advantages than the attackers thus the guard could not be defeated. The guard would chase the nearest attacker and catch them if their distance is no more than the capture distance of the guard, then the caught attacker would be removed. The attacker would move according to the resultant forces as mentioned before while the guard would follow the simplest principle that preference to the location which has a shorter distance to the attacker.

The guard movement rules are defined as follows:
a) Calculate respectively the distances between the guard and every attacker by Eq(2) and find out the location of the nearest attacker.
b) Determine all available positions for the guard, and the available position means the location unoccupied by obstacles, pedestrians or the attacker.
c) Calculate respectively the distance between the nearest attacker and each available position.
d) Move into the location which has the shortest distance to the nearest attacker at the next time step. Note that the guard would select randomly any of them when two or more positions have the same priority and stay unmoved when all neighborhood positions have been occupied.
e) Catch the attacker once the attacker is within the capture distance of the guard.

$$d_{ag} = \sqrt{(x_a - x_g)^2 + (y_a - y_g)^2} \quad (2)$$

where $d_{ag}$ denote the distance between the guard and the attacker. $x_a$ and $y_a$ represent the coordinate values of the attacker. $x_g$ and $y_g$ represent the coordinate values of the guard, respectively.

**2.3 Action of the pedestrian**

The interactions of pedestrians with evacuation scenario, the guard and the attackers must be considered when simulating the movement of people. In this model, pedestrian dynamics is described by the extended floor field model where individuals make their decision according to the so-called transition probabilities modified by several floor fields. We employ Moore neighborhood, composed of a central cell and its eight surrounding cells, as the pedestrian moving method. Therefore, a 3 × 3 matrix of preferences $p_{ij}$, as shown in Fig.3, is constructed which contains the transition probabilities of its neighbors for cell ($i,j$). Note that the transition probabilities mean the individual would prefer an optimal direction of higher fields, rather than move in a direction probabilistically[23].

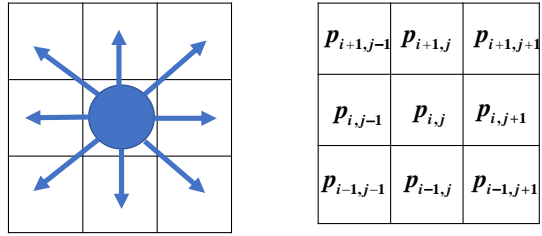

Figure 3 Possible movement directions of the pedestrian and the transition probabilities $p_{ij}$

$$p_{ij} = N \exp\left(\left(k_s S_{ij} + k_t T_{ij} + k_g G_{ij}\right)\left(1 - n_{ij}\right)\right) \qquad (3)$$

Eq(3) describes the transition probabilities $p_{ij}$ of cell($i,j$), where $N$ is a normalization factor. $S_{ij}$, $T_{ij}$, and $G_{ij}$, denote the static floor field, attack threat, and the guard floor field of cell ($i,j$), respectively. And the corresponding sensitivity coefficients are $k_s$, $k_t$, and $k_g$, which determine the weight of each floor field and theoretically range from 0 to 1. If the coefficient is 0, it implies the corresponding floor field has no effect on the pedestrian, and if the coefficient is 0, it means the individual motion is completely determined by the corresponding floor field. In addition, $n_{ij}$ denotes the occupation number of cell (*i,j*) and $n_{ij}$=1 if the cell is occupied by a pedestrian, otherwise $n_{ij}$=0.

In general, the static floor field is related to evacuation scenario such us the size of the room and the location of the exit. In this model, the static floor field $S_{ij}$, calculated by Eq(4), is set inversely proportional to the distance from the exits and could specify regions easily of the room which are more attractive.

$$S_{ij} = N \max_{(i,j)} \left( \min_{e_m} \sqrt{(x_{e_l} - x_{ij})^2 + (y_{e_l} - y_{ij})^2} \right) - \min_{e_m} \sqrt{(x_{e_l} - x_{ij})^2 + (y_{e_l} - y_{ij})^2} \quad (4)$$

where *m* is the number of exit. $e_l$ represent the $l^{th}$ exit. $x_{e_l}$ and $y_{e_l}$ represent coordinate values of the $l^{th}$ exit. $x_{ij}$ and $y_{ij}$ represent coordinate values of cell (*i,j*), respectively.

In contrast, the attack threat, and the guard floor field would evolve with time steps and modified by other individuals. It is certain that the pedestrian would be more willing to try to get away from attackers and prefer a location with long distance to the attacker as possible, thus attack threat $T_{ij}$ is set as Eq(5) to ensure that the location far from the attacker has a higher field and is more attractive to pedestrians. Moreover, some pedestrian might be likely to move close to the guard for survival due to panic when they are chased by the attacker, so the guard floor field $T_{ij}$ can be set as Eq(6) to make sure that the position with short distance from the guard would be more attractive to the pedestrian and has a higher field.

$$T_{ij} = N\sqrt{(x_Q - x_{ij})^2 + (y_Q - y_{ij})^2} \quad (5)$$

where $x_Q$ and $y_Q$ represent coordinate values of the attacker, respectively.

$$G_{ij} = -N\sqrt{(x_g - x_{ij})^2 + (y_g - y_{ij})^2} \quad (6)$$

where $x_g$ and $y_g$ represent coordinate values of the guard, respectively.

## 2.4 Determination of parameter

$k_s$, $k_t$, and $k_g$ are three key parameters reflecting how extent the corresponding factor affect the individual and should be determined before the simulation. Naturally, the sum of these parameters is 1, as Eq(7) shown, because the pedestrian movement is just totally affected by these floor fields in this model.

$$k_s + k_t + k_g = 1 \tag{7}$$

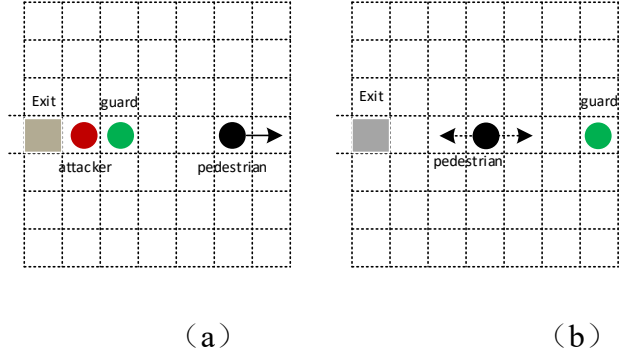

（a） （b）

Figure 4 Diagram of two critical situations. The red circle represents the attacker, the green is the pedestrian, the black is the pedestrian, and the grey is the exit.

Several critical situations are taken into account to study relations between these parameters. As shown in Fig. 4(a), while the attacker, exit and guard in the same $x$ or $y$-axis, the pedestrian would move far away from the attacker due to the threat, rather than head for the exit even though there is the attractive force from the exit and the guard [19]. Therefore, the relation among these parameters can be described in Eq(8).

$$k_t \geq k_s + k_g \tag{8}$$

Another critical situation where a pedestrian is located in the middle of the exit and the guard as shown in Fig. 5(b). In this condition, most people would prefer to the exit for survival while some might be more willing to get close to the guard to avoid the attack due to panic. Actually, the latter would consider high preference of approaching the guard but the former would not, and the two conditions will be studied and compared below.

Combined with Eqs. (7) and (8), these parameters could be obtained as follows:

$$0.5 < k_t \leq 1 \tag{9}$$

$$0 < k_s \leq 0.5 \tag{10}$$

$$0 < k_g \leq 0.5 \tag{11}$$

## 3 Results and Discussion

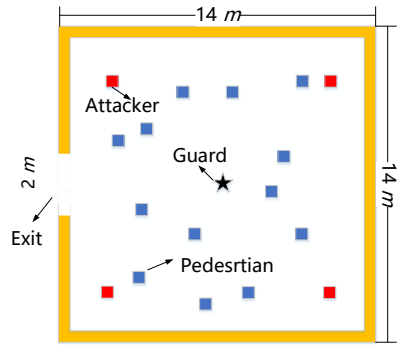

Figure 5 Schematic illustration of the room. Each cell is an identical square area of 0.4 m×0.4 m.

As shown in Figure 5, the proposed model is tested in a cellular space of 35×35 cells and pedestrians are initially distributed randomly and attempt to escape from the room with one exit. And the guard is located in the center of the room at first while the initial positions of attackers are in the corners of the room. The number of the attacker is set to 4 in this senior for a better comparison. When the simulation starts, the attacker would be attracted by pedestrians and move towards the desired direction according to the attractive force. Meanwhile, the guard would chase the nearest attacker, and the pedestrian would escape according to the transition probabilities and would die when assaulted by the attacker. According to the above analysis, the parameters in the simulation are set as follows: $k_t$=0.5, $k_s$=0.4 and $k_g$=0.1 for the situation named "low preference of approaching the guard" where pedestrians would be less affected by the guard and $k_t$=0.5, $k_s$=0.1 and $k_g$=0.4 for the situation named "high preference of approaching the guard" where pedestrian would prefer to move close to the guard for survival. In the following, pedestrian behavior in two strategies and effects of several key parameters on pedestrian flow are investigated.

### 3.1 Effect of the guard on evacuation

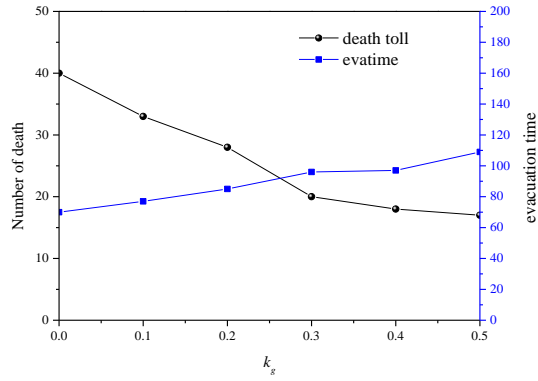

Figure 6 Evacuation time and death toll against $k_g$

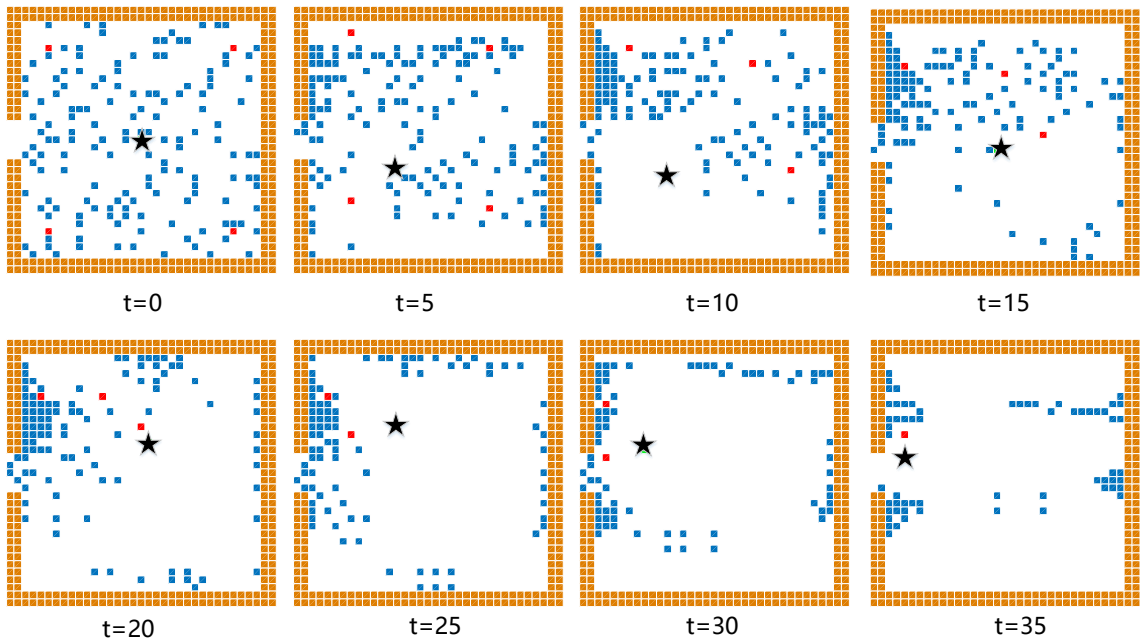

Figure 7 Snapshots at different time steps in the low preference of approaching the guard situation. The black pentagram represents the guard, the red square is the attacker, and the blue is the pedestrian.

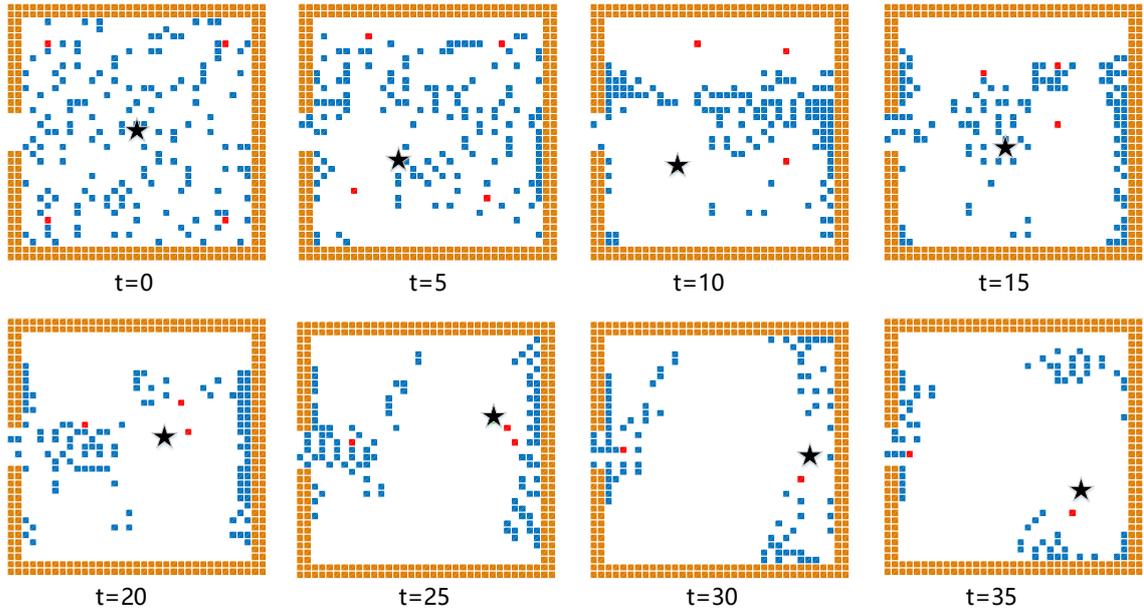

Figure 8 Snapshots at different time steps in the high preference of approaching the guard situation. The black pentagram represents the guard, the red square is the attacker, and the blue is the pedestrian.

Figure 6 indicates how $k_g$ affects the evacuation process. It can be found in Figure 6 that as the $k_g$ -value rises, the numbers of death would decrease but the evacuation time increase. And there are two main reasons for the decrease in death as $k_g$ rises. One is that a larger $k_g$ means the guard would have a greater impact on the decision-making of the individual and the pedestrian would prefer to get close to the guard for survival. Therefore, a certain number of people would stay around the guard to avoid being attacked. Another reason is that the less congestion near the exit reduce the threat of being attacked of the pedestrian. More specially, when $k_g$ is large, many people would be stay around the guard and less congestion near the exit would be generated, and it means the pedestrian has more time to escape the attacker rather than gather around the exit where they are hard to move due to congestion, which would make less death toll. Moreover, the evacuation time increases as $k_g$ rises, and it is because that a larger $k_g$ means the pedestrian would prefer to get close to the guard rather than the exit for survival, which would largely delay the evacuation process.

Further, the specific details are shown in Figure 7, which present snapshots at different time steps when pedestrians are less affected by the guard. As shown in Figure 7, pedestrians have to avoid the attack and try to move to the exit thus a large number of pedestrians are moving toward the exit and the congestion would be formed quickly. For the gathered people, it would be difficult to escape the assault due to the congestion, for example, while $t$=15, and when the attacker reaches the exit and the attackers don't need to spend more time on chase because of the high-density of pedestrians, which means the attackers could launch the attacker every time step and thus more death of pedestrians are caused. Meanwhile, because of the clustered pedestrians near the exit, most of the attackers would move to the exit so that the guard don't need too much extra time to chase another after capturing one attacker, which could greatly reduce the totally catch time. In addition, when all the attackers are caught, the pedestrians could complete the evacuation in a short time because most of the people are around the exit.

The crowd present different dynamic characteristics when the pedestrian is greatly affected by the guard, that is, the pedestrian would prefer to get close to the guard for survival. First, there is still congestion at the exit, but it is significantly reduced, and the overall pedestrian distribution would be more dispersed because most pedestrians would move according to the movement of attackers and guards instead of gathering near the exit. Second, the dispersion of the crowd has also led to the dispersion of attackers due to the fact that the attackers would chase the crowd. Therefore, the guard have to spend more time to catch all attackers. For example, there is only one attacker alive in Figure 7 while t=35 but there are two in Figure 8. It also implies that the attackers would have more time to attack people but it would not cause more death, because there are always a large number of people close to the guard and these people would be less likely to be attacked because the attacker who tries to attacker these people would be caught by the guard. At last, the pedestrians would have to take more time to finish the evacuation when the attack threats have been eliminated by the guard due to the scattered pedestrians.

**3.2 Effect of deterrence radius on evacuation**

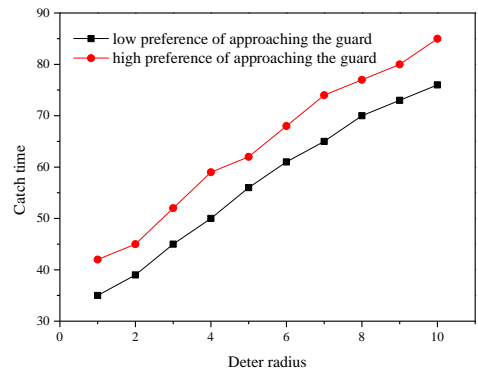

Figure 9 Catch time against the deterrence radius

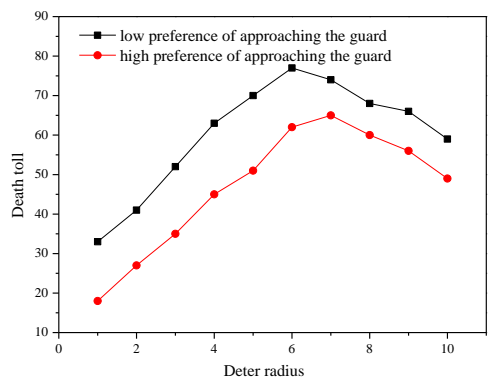

Figure 10 Death toll against the deterrence radius

The deterrence radius represents the critical distance that the attackers have to escape from the guard. Figure 9 presents the catch time against the deterrence radius in two situations. And the catch time is defined as the total time for the guard to successfully capture all the attackers. It can be seen that the catch time shows significant increasing tendency as deterrence radius rises in both cases. When the deterrence radius is large, the attackers would run away from the guard earlier and the guard has to take more time to catch these attackers, which result in the increase in catch time. Moreover, the catch time in the high preference of approaching the guard situation is always higher than that in the low preference situation. In the low preference, the pedestrian would prefer to get close to the exit for survival thus a considerable quantity of pedestrians would stay around the exit, which would make congestion generated near the exit. And the attackers would be attracted by these people and chase them when the distance away from the guard is less than the deterrence radius, and it means the guard who just captured one attacker would take less time to catch another, which leads to the decrease in catch time.

Figure 10 shows how the deterrence radius affects the death toll in the two cases. As the deterrence radius rises, the death toll first increases and then decreases regardless of the low or high preference of approaching the guard. The reason for the increase is that when the deterrence radius is high, the attackers are more difficult to be caught and it would take more time for the guard to grab all attackers, as analyzed before. It also means the attackers have more time to chase and attack people, which would cause more casualties during this process. The reason for the decrease is, if the deterrence radius is large enough, the attackers would largely consider the guard's chase and avoid being caught by the guard as possible. As a result, the attackers prefer to escape the guard earlier and move to the position far away from the guard instead of the position where the attackers could attack more people. Therefore, less pedestrian can be attacked during the escape process of the attacker and the attackers might be forced to move to the corner of the room with a few people, which lead to less death. Moreover, more casualties would occur in the low preference of approaching the guard situation than the high preference situation. This is because in the high preference situation, the pedestrian would prefer to stay around the guard for survival and the attackers would be killed before they could attack the people those stay around the guard. In contrast, in the low preference situation, there would be more congestion and the crowd clustered near the exit is hard to evacuate and the attackers can kill a considerable number of pedestrians in a short time.

**3.3 Effect of capture distance on evacuation**

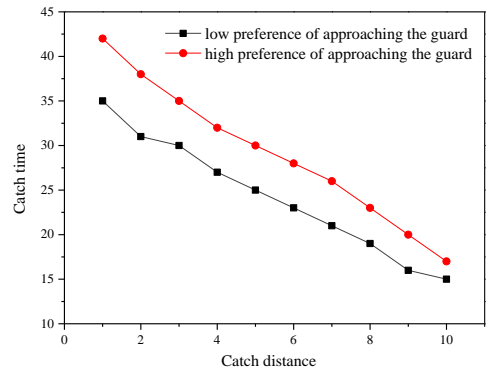

Figure 11 Capture distance against the capture distance

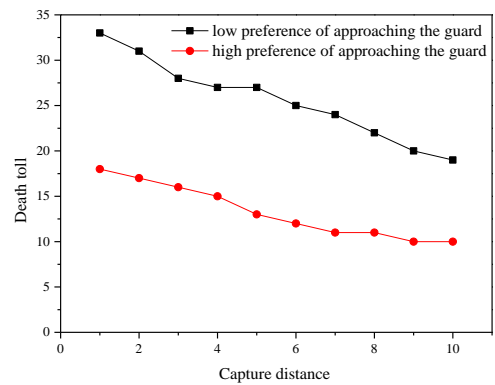

Figure 12 Death toll against the capture distance

The capture distance is the distance that the guard could catch the attackers. Figure 11 indicates the effect of the capture distance on the catch time. It can be seen that the catch time always decreases with the increasing capture distance whether the low or high preference of approaching the guard. When the capture distance is large, the guard could catch the attackers in a long distance thus the catch time would decrease. It should be noted the catch time in the high preference is always higher than that in the low preference. As mentioned before, in the low preference situation, it is more likely for pedestrians to be assembled near the exit due to the tendency of moving towards to the exit. Then the attacker would be attracted by these people and get more close to each other on the distance, and the guard would spend less time to capture these attackers.

Figure 12 shows how the capture distance affects the death toll. It is obvious that less pedestrian would be killed as the capture distance increase. When the capture distance is large, the catch time would decrease due to the fact that the guard could catch the attackers in a long distance. Therefore, the time for the attackers to chase and assault pedestrians would become shorter and the casualties would be reduced. Furthermore, more casualties would occur in the low preference situation. The reason for this is, in the high preference, the pedestrian would prefer to get close the guard for survival, and the attacker would be hard to kill these people because the attacker would be caught before launching the attack. In contrast, in the low preference situation, more congestion would be generated and the attackers don't need much time to chase pedestrians due to the high-density targets near the exit, which means the attackers could kill more people in a quite short time. The two reasons together lead to the decrease in the death toll in the high preference situation.

**4 Conclusion**

This work is aim to study the pedestrian dynamics considering action of the guard in the context of artificial attacks. And two different strategies were compared and the effect of deterrence radius and capture distance on the pedestrian dynamic were studied.

The sensitivity coefficient $k_g$ reflecting the extent of the effect of the guard on the decision-making of pedestrians was investigated. As $k_g$ rises, the pedestrian would be more likely to get close to the guard for survival thus more people would be stay around the guard which would reduce the death of pedestrian. Moreover, when $k_g$ is large, the pedestrian would prefer to the guard instead of the exit for survival, which would largely delay the evacuation process and cause a higher evacuation time.

Two different evacuation strategies named low and high preference of approaching the guard were compared. Compared to the low preference situation, the guard always takes more time to capture all the attackers and less death would be caused in the high preference situation.

The deterrence radius is defined as the critical distance that the attackers have to escape from the guard. As the deterrence radius rises, the attackers would run away from the guard earlier and the guard has to take more time to catch these attackers. Further, the death toll would first increase then decrease. Moreover, as the distance that the guard could catch the attackers rises, it would take much less time for the guard to capture all the attackers.

This study is expected to provide valuable insights to optimize the evacuation strategy. While in reality, the dynamics of pedestrians, the attacker and the guard would be more complex thus certain reasonable simplifications and assumptions have to be made, and some factors such as the attackers' initial distribution and number, which might affect the pedestrian dynamics and would be studied profoundly in the future.

**Appendix: Attacker movement update rules**

Table 1 reflects the attacker movement update rules at each time step for the attacker at $(i, j)$. The attack would select one position which has not been occupied according to the priority position sequence to move. The attacker would select randomly any of them when two positions have the same priority and stay unmoved when all priority positions have been occupied.

Table 1 The attacker movement update rules at each time step for the attacker at (i,j).

| Step 1 | Step 2 | Step 3 | priority position sequence |
|---|---|---|---|
| if $a_t\sin\omega_a>0$ | if $a_t\cos\omega_a>0$ | if $=a_t\sin\omega_a > a_t\cos\omega_a$ | (i-1,j+1)> (i-1,j)> (i,j+1) |
| | | if $=a_t\sin\omega_a = a_t\cos\omega_a$ | (i-1,j+1)> (i-1,j)= (i,j+1) |
| | | if $=a_t\sin\omega_a < a_t\cos\omega_a$ | (i-1,j+1)> (i,j+1) > (i-1,j) |
| | if $a_t\cos\omega_a=0$ | -- | (i-1,j)> (i-1,j+1)= (i-1,j-1) |
| | if $a_t\cos\omega_a<0$ | if $=a_t\sin\omega_a > a_t\cos\omega_a$ | (i-1,j-1) > (i-1,j)> (i,j-1) |
| | | if $=a_t\sin\omega_a = a_t\cos\omega_a$ | (i-1,j-1)> (i-1,j)= (i,j-1) |
| | | if $=a_t\sin\omega_a < a_t\cos\omega_a$ | (i-1,j-1)> (i,j-1) > (i-1,j) |
| if $a_t\sin\omega_a=0$ | if $a_t\cos\omega_a>0$ | -- | (i,j+1)> (i-1,j+1)= (i+1,j+1) |
| | if $a_t\cos\omega_a=0$ | -- | (i,j)= (i,j) |
| | if $a_t\cos\omega_a<0$ | -- | (i,j-1)> (i-1,j-1)= (i+1,j-1) |
| if $a_t\sin\omega_a<0$ | if $a_t\cos\omega_a>0$ | if $=a_t\sin\omega_a > a_t\cos\omega_a$ | (i+1,j+1)> (i+1,j)> (i,j+1) |
| | | if $=a_t\sin\omega_a = a_t\cos\omega_a$ | (i+1,j+1)> (i+1,j)= (i,j+1) |
| | | if $=a_t\sin\omega_a < a_t\cos\omega_a$ | (i+1,j+1)> (i,j+1) > (i+1,j) |
| | if $a_t\cos\omega_a=0$ | -- | (i-1,j+1)> (i-1,j)> (i,j+1) |

| | | if =$a_t\sin\omega_a$ > $a_t\cos\omega_a$ | (i+1,j-1)> (i+1,j)> (i,j-1) |
| --- | --- | --- | --- |
| | if $a_t\cos\omega_a$<0 | if =$a_t\sin\omega_a$ = $a_t\cos\omega_a$ | (i+1,j-1)> (i+1,j)= (i,j-1) |
| | | if =$a_t\sin\omega_a$ < $a_t\cos\omega_a$ | (i+1,j-1) > (i,j-1) > (i+1,j) |